\documentstyle[11pt,twoside,sympo2011,epsfig]{article}
\begin{document}
\title{The 55~Cancri System: Fundamental Stellar Parameters, Habitable Zone Planet, and Super-Earth Diameter}
\author{K. von Braun$^1$, T. S. Boyajian$^{2,3}$, T. A. ten Brummelaar$^6$, G. T. van Belle$^4$, S. R. Kane$^1$, D. R. Ciardi$^1$, M. Lopez-Morales$^5$, H. A. McAlister$^2$, G. Schaefer$^6$, S. T. Ridgway$^7$, L. Sturmann$^6$, J. Sturmann$^6$, R. White$^2$, N. H. Turner$^6$, C. Farrington$^6$, \& P. J. Goldfinger$^6$}
\affil{$^1$NExScI/Caltech, $^2$Georgia State, $^3$Hubble Fellow, $^4$ESO, $^5$CSIC-IEEC/CIW-DTM, $^6$CHARA, $^7$NOAO [kaspar@caltech.edu]} 
\begin{abstract}
The bright star 55~Cancri is known to host five planets, including a transiting super-Earth. We use the CHARA Array to directly determine the following of 55~Cnc's stellar astrophysical parameters: $R=0.943 \pm 0.010 R_{\odot}$, $T_{\rm EFF} = 5196 \pm 24$ K. Planet 55~Cnc f ($M \sin i = 0.155 M_{Jupiter}$) spends the majority of the duration of its elliptical orbit in the circumstellar habitable zone (0.67--1.32 AU) where, with moderate greenhouse heating, it could harbor liquid water. Our determination of 55~Cancri's stellar radius allows for a model-independent calculation of the physical diameter of the transiting super-Earth 55~Cnc e ($\simeq 2.1 R_{\earth}$), which, depending on the assumed literature value of planetary mass, implies a bulk density of 0.76 $\rho_{\earth}$ or 1.07 $\rho_{\earth}$. 
\end{abstract}
\section{Introduction}

55 Cancri (= HD~75732) is a late G / early K dwarf / subgiant currently known to host five extrasolar planets with periods between around 0.7 days and 14 years and minimum masses between 0.026 and 3.84 $M_{Jupiter}$. The super-Earth 55~Cnc e was recently discovered to transit (Winn et al. 2011, Demory et al. 2011), prompting a number of studies of the properties of this system (e.g., Kane et al. 2011, von Braun et al. 2011b). We used the CHARA Interferometric Array to directly measure the stellar angular diameter, which, when combined with Hipparcos parallax measurement and calculation of bolometric flux based on spectral templates and literature broad-band photometry, yields the physical $R_{star}$ and $T_{\rm EFF}$ (von Braun et al. 2011a).

\section{Properties of the 55~Cancri System}

Details of our observations and calculations of stellar properties are described in von Braun et al. (2011b). We give our results in Table 1. Of particular interest are the location and extent of the circumstellar habitable zone (HZ), based on the equations of Jones \& Sleep (2010), and the physical radius of the transiting super-Earth, based on simply the measured flux decrement during transit and our calculated stellar radius value. 

Fig. 1 illustrates that planet f, with its elliptical orbit ($e\simeq0.4$), spends about 74\% of its year inside the HZ, while its equilibrium temperature varies between 221K (apastron) and 302K (periastron) for the assumption of perfect efficiency in the redistribution of energy received from the star (von Braun et al. 2011b).

Coupled with the recent literature values of assumed planetary mass and measured flux decrement during transit, our measured stellar radius implies $R_{p} = 2.007 \pm 0.136 R_{\earth}$ and $\rho_{p} =1.067 \pm 0.132 \rho_{\earth}$ (for $M_{p}$ = $8.63 \pm 0.35 M_{\earth}$; Winn et al. 2011), and $R_{p} = 2.193 \pm 0.146 R_{\earth}$ and $\rho_{p} = 0.757 \pm 0.109 \rho_{\earth}$ (for $M_{p}$ = $7.98 \pm 0.69 M_{\earth}$; Demory et al. 2011).

\begin{table}[h] 
\begin{center}
\begin{tabular}{rc}       
\hline\hline                 
Parameter & Value \\
$\theta_{\rm UD}$ (milliarcseconds) \dotfill		&	$0.685 \pm 0.004$		\\
$\theta_{\rm LD}$ (milliarcseconds)	\dotfill	&	$0.711 \pm 0.004$	\\
Radius ($R_{\rm \odot}$) \dotfill	&	$0.943 \pm 0.010$		\\
Luminosity ($L_{\rm \odot}$) \dotfill	& $0.582 \pm 0.014$		\\
$T_{\rm EFF}$ (K)	\dotfill			&	$5196 \pm 24$			\\
HZ boundaries (AU) \dotfill & $0.67 - 1.32$  \\
\hline\hline
\end{tabular}
\caption{Stellar Properties of the 55~Cancri System. $\theta_{\rm UD}$ and $\theta_{\rm LD}$ correspond to the uniform disk and limb-darkening corrected angular stellar diameters, respectively. $\theta_{\rm LD}$ corresponds to the angular diameter of the Rosseland, or mean, radiating surface of the star.}             
\end{center}
\end{table}

\begin{figure}[h]
\begin{center}
\epsfig{width=7cm,file=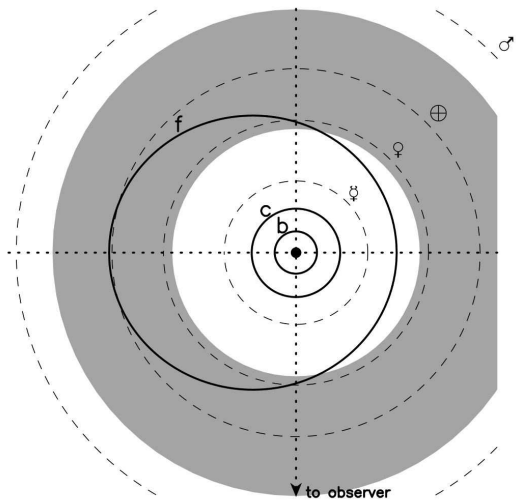}
\caption{Architecture of the 55~Cnc system (planets d and e not shown). Orbital parameters are from Dawson \& Fabrycky (2010). The HZ of the 55~Cnc system is shown as the grey shaded region. Planet f spends approximately 74\% of its year in the HZ. Solar system planet orbits are shown to provide a distance scale, and Mercury's orbital eccentricity is set to zero for clarity.}
\end{center}
\end{figure}



\begin{references}
\reference{  
Dawson, R.~I., \& Fabrycky, D.~C.\ 2010, \apj, 722, 937

Demory, B.-O., et al.\ 2011, arXiv:1105.0415\

Jones, B. W., \& Sleep, P. N. 2010, \mnras, 407, 1259

Kane, S.~R., Gelino, D.~M., Ciardi, D.~R., Dragomir, D., \& von Braun, K.\ 2011, arXiv:1105.1716

von Braun, K., et al.\ 2011a, \apjl, 729, L26

von Braun, K., et al.\ 2011b, arXiv:1106.1152

Winn, J.~N., et al.\ 2011, arXiv:1104.5230
}
\end{references}
\end{document}